\documentstyle[epsf,aps,eqsecnum,multicol]{revtex} 
\def\vec#1{{\rm\bf #1}}

\begin{document}
\title{Strong-Segregation Theory of Bicontinuous Phases in Block
Copolymers} 
\author{Peter D. Olmsted$^{\ast 1}$ and Scott T. Milner$^2$}
\address{$^1$Department of Physics, University of Leeds, Leeds LS2~9JT,
UK and $^2$Exxon Research \& Engineering Company, Corporate Research
Science Laboratories, Annandale NJ 08801}
\date{\today}
\maketitle
\begin{abstract}
We compute phase diagrams for $A_nB_m$ starblock copolymers in the
strong-segregation regime as a function of volume fraction $\phi$,
including bicontinuous phases related to minimal surfaces
(G, D, and P surfaces) as candidate structures. We present the details
of a general method to compute free energies in the strong segregation
limit, and demonstrate that the gyroid G phase is the most nearly stable
among the bicontinuous phases considered. 
We explore some effects of conformational asymmetry on the topology of
the phase diagram.
\end{abstract} 
\begin{multicols}{2}
\narrowtext

\section{Introduction}

Block copolymers (BCPs), comprising chemically distinct polymers
permanently linked together, are interesting because of the diverse
array of ordered phases to which both polymer theory and experiment
have been directed.\cite{bates88,batesglenn90} The phase behavior of
diblock copolymer melts is a competition between the entropic tendency
to mix the two species into an isotropic melt, and an energetic
penalty for having unlike species adjacent, which induces transitions
into ordered phases of many symmetries, depending on the topology and
composition of the polymers. Near the order-disorder transition (weak
incompatibility) entropy effects dominate, and the individual polymers
retain (within mean field) their Gaussian coil conformation through
the transition,\cite{leibler80,note}, while at much higher
incompatibilities the chains are strongly stretched. It is this
strongly stretched regime which we address here.

Leibler developed the first complete theory of ordered phases in BCP
melts\cite{leibler80}, and predicted the by-now classical phases of
lamellar (L), cylindrical (C) and spherical (S) symmetry using the
random phase approximation to derive an effective Landau free energy
in terms of composition modulations in Fourier space. The strong
segregation regime was studied by Helfand and co-workers
\cite{helfand} and Semenov \cite{semenov85}, who predicted the same
series of phases with increasing asymmetry, denoted by the fraction
$\phi$ of polymer A in an $A\!-\!B$ diblock.  (In this work we always
use A to denote the minority block). This treatment balances the
stretching energy of a polymer chain with the interfacial energy
between A and B regions. By assuming an incompressible melt,
minimization of the free energy gives a preferred domain size which
scales as $N^{2/3}$, where $N$ is the degree of polymerization.

In the strong segregation limit the free energies of all microphases
scale the same way with chain length and interfacial tension, so the
phase boundaries become independent of the strength of the repulsion
$\chi$ between A and B monomers and depend only on the composition
$f$. Semenov's calculation in effect gave a lower bound to the free
energy of the L, C, and S phases because the phases he constructed did
not fill space, but were micelles of the corresponding topology
\cite{olmsted94a}. This approximation treats the $A\!-\!B$ interface
and outer block surface as having the same circular or spherical
shape, and is sufficient for understanding the qualitative aspects of
the transitions between the phases.

Experiments followed the theories of Leibler and Semenov and quickly
discovered a new phase,\cite{thomas86,hasegawa87,anderson88}, originally
thought to be ordered bicontinuous double diamond (here denoted D), of
$Pn\bar{3}m$ symmetry, but recently shown to be of $Ia\bar{3}d$ symmetry
\cite{Hajd+94,Fors+94,Schu+96} and related to the minimal surface known as the
gyroid (G).\cite{schoen70} The G phase occurs for compositions between
those of the L and C phases, can occur directly from the disordered
phase upon increasing the incompatibility $\chi N$, and is found
to be unstable to the L or C phases at high enough $\chi N$.\cite{Fors+94} 

Although several groups attempted to describe this transition
theoretically,\cite{MayeDela91,JoneDela94,HamlBate94} using variations
on Leibler's theory, the first successful theory is due to Matsen and
Schick \cite{MatsSchi94e}, who developed a method for computing the
free energy of any crystalline structure by expanding the partition
function in the basis functions for the symmetry of the desired
mesophase, rather than the Fourier mode expansion of Leibler.  They
found a stable gyroid phase for $11.14 < \chi N \alt 60$, where the
upper limit was determined by extrapolation from the phase boundaries
at lower $\chi N$.\cite{MatsBate96} This was followed by careful
application of Leibler's method,\cite{milner97,podneks96} to include
higher harmonics and calculate the stability of the G phase in weak
segregation analytically.

Roughly concurrent to the calculations of Matsen and Schick, methods
were developed to calculate the free energy of essentially arbitrary
structures in the strong segregation regime ($\chi N\rightarrow\infty$).
\cite{olmsted94a,LikhSeme94}. These methods use the results for
polymer brushes,\cite{semenov85,mwc88a}, supplemented by an ansatz
about the geometry of the relevant phase and an assumption about the
chain paths. Olmsted and Milner assumed straight paths through the
$A\!-\!B$ interface and locally specified the volume fraction per
molecule,\cite{olmsted94a,Miln94,Miln94b}, while Likhtman and Semenov
relaxed the assumption of straight paths \cite{LikhSeme94} but enforced
the constraint of constant $\phi$ per molecule only globally. The former
approach corresponds to an upper bound on the free energy (see below),
while it is not clear that the Likhtman-Semenov calculations corresponds
to any bound, or indeed to any systematic approximation, because the
local constraint of constant composition is relaxed. By comparing upper
bounds between bicontinuous, C, and L phases (obtained for the cylindrical
phase by assuming hexagonal symmetry and imposing straight paths), we
showed that the bicontinuous phases are unstable, when comparing upper
bounds, to the L and C phases. Later, Xi and Milner extended this work
to calculations with kinked polymer paths, and found an upper bound to
the hexagonal phase which lies very close to the lower bound using round
unit cells.\cite{XiMiln96b}

Experiments have found an additional phase at $\chi$ values between
the G and L phases \cite{hamley93}, a hexagonally-perforated lamellae
(HPL) phase, which consists of majority lamellae connected through a
minority matrix by hexagonal arrays of tubes.\cite{disko93} The
stacking has been suggested to be $ABCABC$ \cite{Fors+94} or $ABAB$
\cite{hamley93}. Theoretical attempts to justify this phase have
failed in both the strong segregation limit, where Fredrickson chose a
catenoid as a candidate base surface;\cite{fredrickson91} and in the
weak-segregation limit by self-consistent field calculations
\cite{MatsBate96}. Recent experiments \cite{hajdukunpub} have shown
that the HPL phase is not an equilibrium phase in diblock melts, but
may be metastable.

Here we present the calculations of Ref.~\cite{olmsted94a} in more
detail. We show that the G geometry is the most stable of the
candidate bicontinuous phases, followed by the D and P geometries, and
that the G phase can be stable for block-copolymers with sufficient
conformational asymmetry.  The outline of this paper is as follows. In
Section~\ref{sec:general} we present the formalism for calculating the
free energy in general geometries. In Section~\ref{sec:classical} we
present the results for the classical diblock topologies (lamellae,
cylinders, and spheres), extended to include non-round unit cells and,
in the case of the cylindrical topology, kinked paths. In
Section~\ref{sec:bicontinuous} we present the free energy for a
generic ``saddle'' wedge, which is representative of a generic
bicontinuous structure as a pie-shaped wedge is representative of
cylindrical phases regardless of packing. We then introduce the
geometry necessary for calculating the free energy of the P, D, and G
topologies.  In Section~\ref{sec:results} we present our results for
both symmetric and non-symmetric stars, and we conclude in
Section~\ref{sec:summary}.

\section{Strong Segregation Theory in general geometries}\label{sec:general}
\subsection{A single wedge} \label{sec:2a}
We first recall some results for polymer brushes in strong segregation
under melt conditions, and then show how to apply this to a general
geometry. We consider a melt of star $A_{n_{\scriptscriptstyle
    A}}\!-\!B_{n_{\scriptscriptstyle B}}$ copolymers , comprising
$n_{\scriptscriptstyle A}$ arms of \textit{A}-blocks of mean square end-to-end
distance $ R_{\scriptscriptstyle A}$ and similarly for the \textit{B}-arms.
The volume fraction $\phi$ of \textit{A} material is \cite{mmnote}
\begin{equation}
\phi = {n_{\scriptscriptstyle A} \Omega_{\scriptscriptstyle A}\over
\Omega}, 
\end{equation} 
where $\Omega=n_{\scriptscriptstyle A}\Omega_{\scriptscriptstyle
  A}+n_{\scriptscriptstyle B}\Omega_{\scriptscriptstyle B}$ is the
total chain volume, and $\Omega_{\scriptscriptstyle A}$ and
$\Omega_{\scriptscriptstyle B}$ are the volumes of single \textit{A}
and \textit{B} arms. Our calculations are appropriate for
strongly-\-seg\-re\-ga\-ted chains, for which interfaces are sharp on
the scale of microphase lattice constants. In strong segregation the
free energies of all microphases scale the same way with chain length
and interfacial tension, so the phase boundaries become independent of
the strength of the repulsion $\chi$ between \textit{A} and \textit{B}
monomers.
{\begin{figure}
\displaywidth\columnwidth
\epsfxsize=2.5truein 
\centerline{\epsfbox{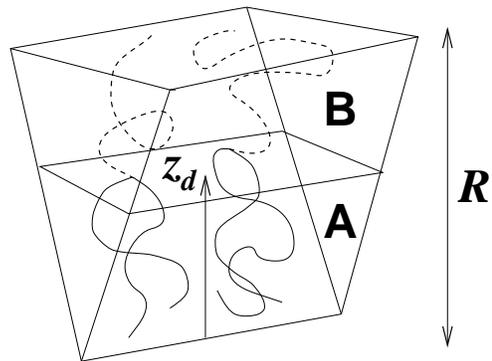}}  
\caption{
A typical wedge filled with an $A\!-\!B$ star copolymer, 
with $n_{\scriptscriptstyle A}=2$ $A$ arms and $n_{\scriptscriptstyle
B}=1$
$B$ arm.}
\label{fig:wedge}
\end{figure}}

Consider an elementary wedge, as in Figure \ref{fig:wedge}, from which we
will construct all of the strong-segregation phases. Our calculations
are performed in terms of the ratio 
\begin{equation}
a(\lambda)\equiv {{\cal A\/}(z)\over {\cal A\/}(R)}\qquad(\lambda
=z/R) \label{eq:area}
\end{equation} 
of the cross-sectional area ${\cal A\/}$ at a height $z$ relative to
that of the outer surface, in an infinitesimal wedge of height $R$.
This function may be easily calculated for wedges of particular shapes
by elementary geometry, and is collected in Table~\ref{table1} for
various geometries. Since $a(\lambda)$ is the projected surface area
along the normal vector extending from the wedge point to the flat
wedge top, it will be a quadratic function of $\lambda$. The boundary
condition $a(0)=0$ implies that $a(\lambda)$ is a sum of $\lambda$ and
$\lambda^2$ terms, and the boundary condition $a(1)=1$ fixes the sum
of the corresponding coefficients to be $1$, leaving a single
parameter. Hence we may generally write
\begin{equation}
a(\lambda) = p \lambda + (1-p) \lambda^2.
\label{eq:p}
\end{equation}

The location $z_d$ of the ``dividing surface'' separating the
two species is determined by equating the relative volume below
$z$, denoted $v(z/R)$, to the volume fraction $\phi$:
\begin{eqnarray}
v(z/R)&\equiv&\int_0^{z/R}\!dy\,a(y)
= {V(z)\over R {\cal A\/}(R)}\label{eq:vdef} \\
v(\beta) &=&\phi \, v(1) \quad \qquad (\beta=z_d/R) \quad , \label{eq:zdloc} 
\end{eqnarray}
where $V(z)$ is the (partial) volume of the wedge below height $z$ (see
Figure \ref{fig:wedge}).
\begin{table}[!htb]  
\caption{Area function $a(\lambda)$ (see eq~\ref{eq:area}) for  
various wedge geometries, where $\lambda=z/R$ and $R$ is the wedge height.  
The expression for $p$ for the D, P, and G wedges refers to the geometry  
in Figure~\ref{fig:PGwedges}.}  
\begin{tabular}{lc}  
Structure&$a(\lambda)$\\  
\hline  
Lamellae & $1$ \\  
Cylinders $(p=1)$& $\lambda$ \\  
Spheres $(p=0)$& $\lambda^2$ \\  
Symmetric Wedge (Figure~\ref{fig:chip}) $(p=2)$ & $\lambda(2 - \lambda)$ \\  
D, P, G Wedges (Figure~\ref{fig:PGwedges}) & $p \, {\lambda} + (1 -  
p) \lambda^2$ \\  
\end{tabular}  
\tablenotetext{$p={\bf \hat{n}}\cdot  
(\vec{z}_1\!-\!\vec{z}_2)\times(\vec{d}\!-\!\vec{b}+  
\vec{a}\!-\!\vec{c})/ 2 {\bf \hat{n}}\cdot   
(\vec{a}\!-\!\vec{c})\times(\vec{d}\!-\!\vec{b})$\hfill(D,P)}   
\tablenotetext{$\phantom{p=}\,\,{\bf \hat{n}}\cdot  
(\vec{z}_1\!-\!\vec{z}_2)\times(\vec{d}\!-\!\vec{b})/  
2 {\bf \hat{n}}\cdot   
(\vec{a}\!-\!\vec{c})\times(\vec{d}\!-\!\vec{b})$\hfill(G)}  
\label{table1}  
\end{table}  

{\begin{figure}
\displaywidth\columnwidth
\epsfxsize=3.0truein 
\centerline{\epsfbox{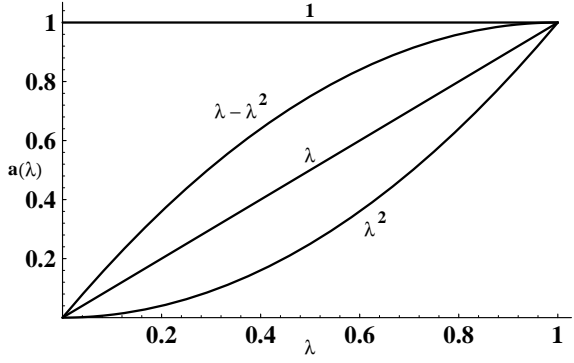}}  
\caption{ Area functions $a(\lambda)= 1, \lambda - \lambda^2, \lambda,$
and
$\lambda^2$, for the lamellar, symmetric wedge, cylinder and sphere
phases. }
\label{fig:area}
\end{figure}}

The free energy per molecule of a wedge in the strong segregation limit is
the sum of stretching and interfacial contributions,
\begin{equation}
f = f_{\mit int\/} + f_{\mit str\/}.
\label{eq:ftot}
\end{equation}
The interfacial free energy per molecule is simply the surface tension
contribution which, per chain, is
\begin{equation}
f_{int}= {\gamma \,\Omega \, a(\beta)\, \over v(1)\, R}
\label{eq:interf}
\end{equation}
where the surface tension $\gamma$ 
scales as $\chi^{1/2}$.\cite{helfand71} In the strong segregation limit
we ignore the translational entropy of the junction points, which scales
logarithmically with molecular weight and is thus subdominant.

The stretching energy is calculated by methods developed for polymer
brushes,\cite{semenov85,mwc88a}. The copolymer chains are added
one by one, and the work to add each is summed. The height of the layer
$h_{\scriptscriptstyle A}$ when the number of chains per area is $\sigma
(h_{\scriptscriptstyle A})$ is given by
\begin{eqnarray}
v(\beta) - v(\beta-h_{\scriptscriptstyle A}/R)&=&
\sigma(h_{\scriptscriptstyle A}) n_{\scriptscriptstyle A} 
\Omega_{\scriptscriptstyle A} a(\beta) / R \label{eq:htcov0} \\
v(\beta+h_{\scriptscriptstyle B}/R)-v(\beta)&=&
\sigma(h_{\scriptscriptstyle B}) n_{\scriptscriptstyle B} 
\Omega_{\scriptscriptstyle B} a(\beta) / R
\label{eq:htcov}
\end{eqnarray}
where $h_{\scriptscriptstyle A}$ is the height of the growing A-layer, which
is measured relative to the junction at $z_d$; and similarly for $B$.

An important quantity is the monomer chemical potential $\mu(\phi)$
(the hydrostatic pressure, in an incompressible system), which is a
decreasing function of distance from the dividing surface and is
microscopically responsible for the stretching of the chains as their
monomers seek regions of lower chemical potential. Under the
assumption that there are free ends at all distances from the dividing
surface, the chemical potential is quadratic in the distance
from the dividing surface:\cite{mwc88a}
\begin{equation}
\mu(z_{\scriptscriptstyle A}) = 
{3\pi^2\over 8 \Omega_{\scriptscriptstyle A}
 R_{\scriptscriptstyle A}^2}\left[
h_{\scriptscriptstyle A}^2
- \left(z_{\scriptscriptstyle A} - z_d\right)^2\right].
\label{eq:mu}
\end{equation} 
A similar equation holds for the $B$-species.
For the inwardly curved parts of the
structure, this is exact; for the outwardly curved parts, this
assumption leads to an unphysical negative density of free
ends \cite{semenov85}, but has been shown to give extremely good
estimates of stretching free energy even for layers with curvature
radii comparable to their thickness.\cite{ballbrush91}

The work to add an A-block is independent of the location of the free
end, and so may be conveniently taken to be the work to add a chain
with its conformation very near the surface, simply 
$\Omega_{\scriptscriptstyle A}
\mu(z_d)$. Hence the total free energy of a chain is obtained
by integrating up to the desired coverage $\sigma$,
\begin{equation}
f_{\mit str} = {3\pi^2\over 8\sigma} 
\int_0^{\sigma}d{\sigma}' \left[
{n_{\scriptscriptstyle A}\over  R_{\scriptscriptstyle A}^2}h_{\mit
\scriptscriptstyle A}^2\left({\sigma}'\right)
+ {n_{\scriptscriptstyle B}\over  R_{\scriptscriptstyle B}^2}h_{\mit
\scriptscriptstyle B}^2\left({\sigma}'\right)\right]. 
\label{eq:fstr0}
\end{equation}
Using eqs~\ref{eq:htcov0}-\ref{eq:htcov} we can write
\begin{eqnarray}
d\sigma &=& {dh_{\scriptscriptstyle A}\over\Omega \phi}\,
{a(\beta + h_{\scriptscriptstyle A}/R)\over a(\beta)}, \\
d\sigma &=& {dh_{\scriptscriptstyle B}\over\Omega (1-\phi)}\,
{a(\beta - h_{\scriptscriptstyle B}/R)\over a(\beta)}.
\end{eqnarray}
Changing variables from $\sigma$ to $h_{\scriptscriptstyle A}$ 
and $h_{\scriptscriptstyle B}$ and using
eqs~\ref{eq:htcov0}-\ref{eq:htcov}, we rewrite eq~\ref{eq:fstr0}
as
\begin{equation}
f_{\mit str}={\pi^2 R^2 \over 8 v(1)} \left[
{n_{\scriptscriptstyle A} \over \phi R_{\scriptscriptstyle A}^2}
I_{\scriptscriptstyle A}
 +
{n_{\scriptscriptstyle B} \over (1-\phi) R_{\scriptscriptstyle B}^2}
I_{\scriptscriptstyle B}\right] 
\label{eq:fstr} 
\end{equation}
where
\begin{eqnarray}
I_{\scriptscriptstyle A} &=& 3 \int_0^\beta\!\!\!dy\, a(\beta-y) y^2 
\label{eq:IA}\\
I_{\scriptscriptstyle B} &=& 3 \int_0^{1-\beta}\!\!\!dy\, a(y+\beta) y^2\;.
\label{eq:IB}
\end{eqnarray}
At this point we make contact with previous calculations of
asymmetric block copolymers and 
introduce an asymmetry parameter $\varepsilon$:\cite{Miln94b,mmnote}
\begin{equation}
\varepsilon^2 = 
{n_{\scriptscriptstyle B}^2\over n_{\scriptscriptstyle A}^2}
{\Omega_{\scriptscriptstyle B}\over R_{\scriptscriptstyle B}^2} { R_{\scriptscriptstyle A}^2\over\Omega_{\scriptscriptstyle A}}
= {n_{\scriptscriptstyle B}^2\over n_{\scriptscriptstyle A}^2}
\varepsilon_{\scriptscriptstyle F}.
\label{eq:asym}
\end{equation}
The second equality relates $\varepsilon^2$ to Fredrickson's
asymmetry parameter $\varepsilon_{\scriptscriptstyle F}$ {\sl
e.g.} Ref.~\cite{BateFred94}). The ratio $\ell_{\scriptscriptstyle A}
=\Omega_{\scriptscriptstyle A}/ R_{\scriptscriptstyle A}^2$ is 
a characteristic length which is
independent of the length of an $A$-arm, and is larger for more flexible
chains at a given volume. A smaller $\varepsilon$ indicates an enhanced
tendency for the $B$ species to stretch. Table~\ref{table2} shows
asymmetry parameters for several diblocks.
\begin{table}[!htb]  
\caption{Asymmetry parameters for diblock copolymers
  $(n_{\scriptscriptstyle A}=n_{\scriptscriptstyle B}=1)$ as
  defined by eq~\ref{eq:asym}.}
\begin{tabular}{lccc}  
A-B&$\varepsilon = \sqrt{{\Omega_{\scriptscriptstyle B}
\over R_{\scriptscriptstyle B}^2} 
{ R_{\scriptscriptstyle A}^2\over\Omega_{\scriptscriptstyle A}}}$
&$\varepsilon_F=\varepsilon^2$&ref\\  
\hline  
PE-PEP& 1.22 & 1.5& \cite{faraday}\\  
PE-PEE& 1.58 & 2.5& \cite{faraday}\\  
PEP-PEE& 1.22 & 1.5& \cite{faraday}\\  
& 1.27 & 1.61& \cite{fetters94}\\  
PI-PS& 1.22 & 1.5& \cite{faraday}\\  
& 1.11 & 1.23& \cite{fetters94}\\  
PB-PI& 1.18 & 1.39& \cite{fetters94}\\  
PB-PS& 1.31 & 1.72& \cite{fetters94}\\  
\end{tabular}  
\tablenotetext{PE=polyethylene, PS=1,4-polystyrene, PI=polyisoprene,
PB=1,4-polybutadiene, PEP=\textit{alt}-poly(ethylenepropylene), 
PEE=poly(ethylethylene)
}
\label{table2}  
\end{table}  

Upon minimizing eq~\ref{eq:ftot} over the scale of the structure,
{\sl i.e.\/} the radius $R$, it is convenient to normalize all energies
by a characteristic energy $f_0$ to obtain a compact form for the
free energy:
\begin{equation}
f = f_0 \left\{{a(\beta)^2\over v(1)^3} \left(
{I_{\scriptscriptstyle A}\over \varepsilon\phi^2} + {\varepsilon 
I_{\scriptscriptstyle B} \over (1-\phi)^2} \right)\right\}^{1/3},
\end{equation}
where \cite{mmnote}
\begin{equation}
f_0 = \left({27 \pi^2\over 32}\right)^{1/3} \Omega^{1/3} \gamma^{2/3} 
(n_{\scriptscriptstyle A} n_{\scriptscriptstyle B})^{1/3}
\left(\ell_{\scriptscriptstyle A} \ell_{\scriptscriptstyle B}\right)^{1/6}
\label{eq:f0}
\end{equation}
is the free energy of a symmetric $(\varepsilon=1)$ lamellar phase.

\subsection{Composite structures}
The procedure above applies to a single wedge. For the classical cylinder and
sphere geometries with circular (spherical) unit cells, each wedge is
identical. For non-circular unit cells and for the complex geometries
of bicontinuous phases, we must assemble the structure
from many different wedges and minimize over the scale factor for the
entire structure. The average free energy per chain $f_{tot}$ of a structure
with many distinct wedges is 
\begin{equation}
f_{tot} = {\sum_{\alpha} f_{\alpha} dV_{\alpha}\over V}, \label{eq:fsum}
\end{equation}
where $f_{\alpha}$ is the free energy per molecule in
wedge $\alpha$ of volume $dV_{\alpha}$, and 
\begin{equation}
V = \sum_{\alpha} dV_{\alpha}
\end{equation}
is the volume of the structure.  We choose a single scale factor $R_0$
to determine the size of the whole structure.  Each wedge $\alpha$ has
its own area and volume functions $a_{\alpha}(\beta_{\alpha})$ and
$v_{\alpha}(\beta_{\alpha})$, where the position of the dividing
surface of each wedge, $\beta_{\alpha}\equiv z_{d\alpha}/R_{\alpha}$,
is determined by
\begin{equation}
v_{\alpha}(\beta_{\alpha}) = \phi v_{\alpha}(1). \label{eq:zdloc2}
\end{equation}

The dimensionless functions $\{a_{\alpha}\}$ are generalizations of
eq~\ref{eq:area} for each wedge $\alpha$ with wedge height
$R_{\alpha}$. These functions encode the geometry of the particular
structure, and the cross-sectional area at the top of each wedge,
$A_{\alpha}(R_{\alpha})$, scales as $R_0^{d-1}$, for a $d$-dimensional
structure ({\sl e.g.\/} $d\!=\!1$ for lamellae, $d\!=\!2$ for
cylinders, $d\!=\!3$ for spheres). Expressing
eq~\ref{eq:interf},\ref{eq:fstr} in terms of $\{a_{\alpha},
A_{\alpha}\}$ and $R_0$, we minimize over $R_0$ to find the following
free energy per chain of a particular structure:
\begin{equation}
f = f_0 \left\{S_a^2 \left(
{S_{\scriptscriptstyle A}\over \varepsilon\phi^2} + {\varepsilon
S_{\scriptscriptstyle B} \over (1-\phi)^2}
\right)\right\}^{1/3}, \label{eq:avg}
\end{equation}
where
\begin{eqnarray}
S_{a} &=& {{\displaystyle \sum_{\alpha} 
A_{\alpha}(R_{\alpha}) a_{\alpha}(\beta_{\alpha})}
\over {\displaystyle \sum_{\alpha} 
v_{\alpha}(1)} R_{\alpha} A_{\alpha}(R_{\alpha})}
\label{eq:Sa} \\
\noalign{\vskip7pt}
S_{\scriptscriptstyle A} &=& {{\displaystyle \sum_{\alpha} 
R_{\alpha}^3 A_{\alpha}(R_{\alpha})
I_{\scriptscriptstyle A}(\alpha,\beta_{\alpha})} \over
{\displaystyle \sum_{\alpha} 
v_{\alpha}(1)} R_{\alpha} A_{\alpha}(R_{\alpha})}, \label{eq:SA}
\end{eqnarray}
where $I_{\scriptscriptstyle A}(\alpha,\beta_{\alpha})$ is obtained
from eqs~\ref{eq:IA}-\ref{eq:IB} by substituting
$a_{\alpha}(\beta_{\alpha})$ in place of $a(\beta)$, and a similar
relation defines $S_{\scriptscriptstyle B}$.

By specifying the volume fraction in each wedge according to
eq~\ref{eq:zdloc2}, we locally satisfy the constraint arising from the
fixed composition of the copolymers. In contrast, Likhtman and Semenov
\cite{LikhSeme94} satisfied this constraint only globally within a
particular structure, which would be relevant for mixtures of different
diblock copolymers with overall composition $\phi$ \cite{ARB90} in the
strong segregation limit, in which the entropy of mixing of such different
copolymers would be negligible

\section{Free energies of classical diblock topologies}\label{sec:classical}
\subsection{Round unit cells}

Using the results of Section~\ref{sec:2a} we can find the energies
of the classical phases of diblock-copolymers: lamellae (L), cylinders
(C), and spheres (S) in the round unit cell approximation, in which
the unit cells are taken to consist of identical wedges. The
corresponding free energies are:
\begin{eqnarray}
{f_{lam} \over f_0} &=& \left[\varepsilon(1-\phi) +
{\phi\over\varepsilon}\right]^{1/3} \label{eq:felam}\\
{f_{cyl} \over f_0} &=& \left[ 
{2 \varepsilon\phi(1-\phi^{1/2})^3(3+\phi^{1/2})
\over (1-\phi)^2} + {2 \phi \over \varepsilon}\right]^{1/3} \label{eq:fecyl} \\
{f_{sph} \over f_0} &=& 3\left[ {\varepsilon\phi^{4/3}(1-\phi^{1/3})^3
(\phi^{2/3}+3\phi^{1/3}+6) \over 10(1-\phi)^2} +
{\phi \over 10\varepsilon} \right]^{1/3}\!\!\!\!\!\!\!. \label{eq:fesph}
\end{eqnarray}
	
Calculations based on round unit cells \cite{semenov85} provide lower
bounds for the free energy, because they in fact describe the free
energy per molecule of micelles.\cite{olmsted94a} We
may imagine a volume packed with such micelles, the interstitial
regions filled with compatible long homopolymer with negligible
surface tension against the outsides of the micelles, and negligible
entropy of mixing. Then we could do work to deform the micelles
into a space-filling array, expelling the homopolymer at no free
energy cost. To distinguish between crystal structures within a
particular topology, such as between hexagonal and square for the
cylindrical topologies, we must examine the energy for packing the
molecules into the particular geometry, which is performed below.
{\begin{figure}
\displaywidth\columnwidth
\epsfxsize=3.0truein 
\centerline{\epsfbox{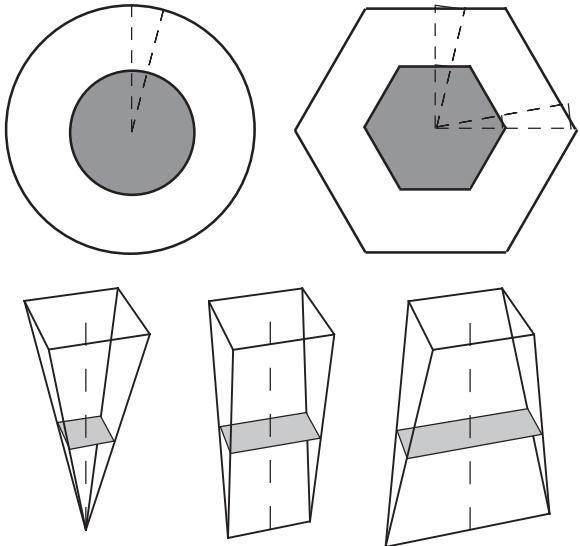}}  
\caption{ (Above) Hexagonal and round unit cells for the cylindrical
phase, and (below) typical wedges for spherical, cylindrical, and
bicontinuous phases; the dividing surface is shaded. }
\label{fig:wedge2}
\end{figure}}

\subsection{Non-round unit cells (straight paths)}

We can produce an {\sl upper} bound for different structures by assembling
small pieces of the cylindrical or spherical micelles to fill the
appropriate unit cell. Each wedge has a parabolic monomer chemical
potential given by eq~\ref{eq:mu}. However, each wedge $\alpha$
has a slightly different shape and geometry, and thus has a distinct
potential $\mu_{\alpha}$. Adjacent wedges are not in equilibrium with each
other and  will relax if allowed to do so. Hence the calculation
yields an upper bound. To construct the unit cell of, {\sl e.g.\/},
hexagonal cylinders, we assume a hexagonal dividing surface scaled
down by $\phi^{1/2}$ and assemble the unit cell from tiny pie-shaped
wedges extending from the center of the hexagon to the cell boundary.
We make an analogous construction for square arrays of cylinders, or
for FCC and BCC packings of spheres.

We calculate the volume-averaged stretching free energy per molecule using
eq~\ref{eq:avg}. To calculate this in practice we use the following
procedure. For cylindrical micelles we divide a cell of a given symmetry
(say, hexagonal) into tiny wedges. Each wedge is adjusted slightly
by making the segment of the wedge on the dividing surface normal to
the bisector of the wedge, which is the path of the polymer. Such an
adjustment introduces a negligible volume in the continuum limit of many
small wedges. The surface area used for calculating the surface energy
(eq~\ref{eq:Sa}) is, of course, the area of the segment in the original
hexagonal dividing surface (before adjusting the wedge to account for
straight paths).

{\begin{figure}
\displaywidth\columnwidth
\epsfxsize=3.5truein 
\centerline{\epsfbox{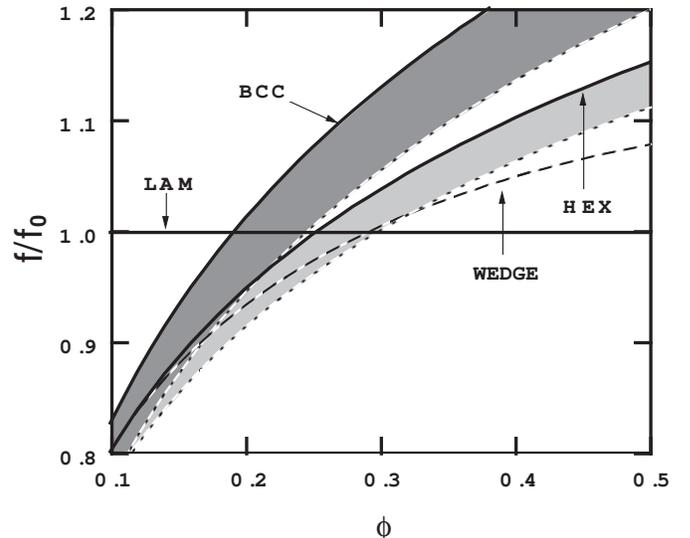}}  
\caption{
Free energies of classical phases and the symmetric wedge,
for straight-paths constructions (asymmetry factor $\varepsilon=1$).
The shaded regions are bounded by the
upper (BCC and HEX, respectively) and lower (spherical and cylindrical
micelles,
respectively) free energy bounds for the spherical and cylindrical
topologies. The symmetric wedge has area
function $a(\lambda) = 2(\lambda-\lambda^2)$, described in
Sec.~\ref{sec:wedge}. }
\label{fig:classicala}
\end{figure}}
For the classical phases (see Figure~\ref{fig:classicala}) the ratios
of the upper and lower bounds for the free energies are independent of
$\phi$, given by:

\begin{equation}
\begin{array}{lcccr}
\displaystyle {f_{\mit hex}\over f_{\mit cyl}} 
&=& {\displaystyle\left({10\over 9}\right)^{1/3}} &\simeq& 1.036 \\
\noalign{\vskip10truept}
\displaystyle {f_{\mit square}\over f_{\mit cyl}} 
&=& {\displaystyle\left({4\over 3}\right)^{1/3}}&\simeq& 1.101 \\
\noalign{\vskip10truept}
\displaystyle {f_{\mit bcc}\over f_{\mit sph}} 
&=& {\displaystyle\left(95\over 384\right)^{1/3}\left({1\over 2}
+\sqrt{3}\right)^{2/3}} &\simeq& 1.072 \\
\noalign{\vskip10truept}
\displaystyle {f_{\mit fcc}\over f_{\mit sph}} 
&=& {\displaystyle\left({5\over 4}\right)^{1/3}} &\simeq& 1.077.
\end{array}
\end{equation}


Evidently, the most favorable structures have the ``roundest'' unit
cells. The hexagonal phase is favored over the square phase, and BCC
is slightly favored over FCC.

\subsection{Non-round unit cells (kinked paths)}

{\begin{figure}
\displaywidth\columnwidth
\epsfxsize=2.0truein 
\centerline{\epsfbox{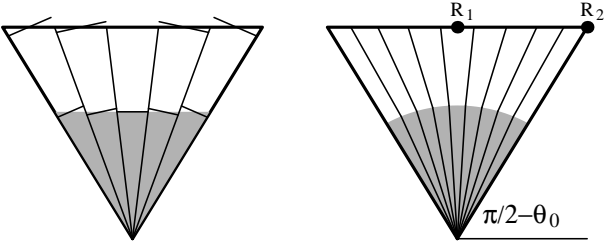}}  
\caption{Hexagonal unit volume divided into wedges for straight-path
(left) and kinked-path (right) ansatzes, for equal compositions
$\phi$. Also indicated are adjustments necessary to calculate the free
energy in each wedge for the straight-path ansatz.}
\label{fig:wedgecut}
\end{figure}}
Other upper bounds can be obtained by using different prescriptions 
for the $A\!-\!B$ surface.
For example, one could choose a circular $A\!-\!B$ surface of radius
$D(\phi)$ for the cylindrical phase. The inner ($A$) volume may be
divided into wedges, and the outer ($B$) volume divided into wedges
which each satisfy the volume constraint $\phi$ with a partner $A$ wedge
\cite{XiMiln96b}. For a right triangular wedge which subtends an angle
$\theta_0$, points at angle $\theta$ on the $A\!-\!B$ surface map to
points $\vec{R}(\theta)$ on the boundary of the Wigner-Seitz cell,
\begin{equation}
\vec{R}(\theta) = \vec{R}_1 + s\!\left(\theta\right) 
\left(\vec{R}_2 - \vec{R}_1\right)
\end{equation}
where $\vec{R}_1=\{0,1\}$, and
$\vec{R}_2=\{\tan\theta_0,1\}$. The composition specifies the radius,
according to 
\begin{equation}
\phi\tan\theta_0=D^2(\phi)\theta_0.
\end{equation}
The mapping which obeys the local composition constraint is
\begin{equation}
s(\theta) = {\displaystyle{\theta\over\theta_0} - D(\phi)
{\sin\theta\over\tan\theta_0}\over 1-D(\phi)\cos\theta},
\end{equation}
where $\theta_0=\pi/3$ and $\pi/4$ for hexagons and squares (see
Figure \ref{fig:wedgecut}),
respectively. 

Minimizing over the scale of the structure, eq~\ref{eq:fsum}
yields, after some calculation, the following free energy:
\begin{eqnarray}
{f\over f_0} &=& \left\{ {2\phi\over 3}\left[ \varepsilon^{-1}
+ {\varepsilon \phi^2\over (1-\phi)^2 D^4(\phi)} \right.\right.\times\\
&&\left.\left.\int_0^{\theta_0} {d\theta\over\theta_0}
|\vec{R}-\vec{r}|^3\left(
3\left|{d\vec{R}'\over d\theta}\right| + \left|{d\vec{r}'
\over d\theta}\right|\right)\right]\right\}^{1/3},
\end{eqnarray}
where 
\begin{eqnarray}
\vec{r}(\theta) &=& D(\phi) {\sin\theta\choose\cos\theta} \\
\vec{r}'&=& \vec{r}\cdot\left[ \bbox{\delta} - 
{(\vec{R}-\vec{r}) (\vec{R}-\vec{r})\over |\vec{R} - \vec{r}|^2}
\right]
\end{eqnarray}
and similarly for $d\vec{R}'$. Remarkably, the upper bound for the
kinked-path-hexagonal ansatz is typically less than $1\%$ above the
lower bound of cylindrical micelles, and the transition is shifted to
only a slightly smaller $A$ fraction $\phi$ (see
Figure~\ref{fig:classicalb}). Apparently the extra stretching energy to
maintain a hexagonal $A\!-\!B$ interface with straight paths is
relaxed considerably by allowing the inner block to adopt a more
nearly circular dividing surface, which is preferred. Recent accurate
numerical self-consistent field calculations of diblock melts have
shown that in fact the $A\!-\!B$ interface is nearly circular, with a
slight hexagonal modulation (angular modulation with 6-fold symmetry)
of relative amplitude $0.03\%$ at $\chi N= 60$ and $\phi=0.33$
\cite{matsen97}.
{\begin{figure}
\displaywidth\columnwidth
\epsfxsize=3.5truein 
\centerline{\epsfbox{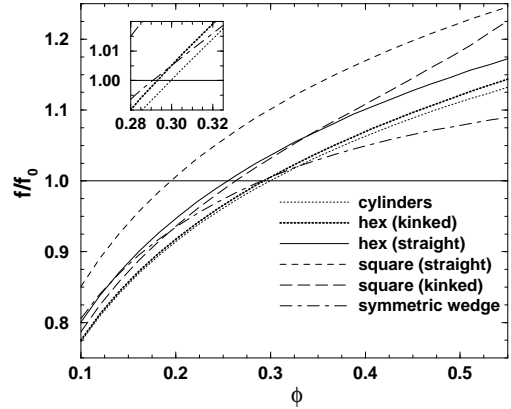}}  
\caption{Free energies for
cylindrical topologies in various approximations (asymmetry factor
($\varepsilon=1$). Also shown is the free
energy of the symmetric wedge. The inset shows an enlarged view of the
crossing of the wedge and kinked-path-hexagonal free energies. The
lamellar energy crosses the cylindrical micelle energy at $\phi=0.299$,
the kinked-path-hexagonal energy at
$\phi= 0.293$, and the straight-path-hexagonal energy at $\phi=0.255$.}
\label{fig:classicalb}
\end{figure}}

\section{Geometry of bicontinuous phases}\label{sec:bicontinuous}
\subsection{Generic saddle surfaces}\label{sec:wedge}
Before addressing particular symmetries (P, D, or G) of bicontinuous
phases, we discuss the closest analogue to a round unit cell. We would
like to produce a simple estimate of the free energy, analogous to the
cylindrical and spherical micelle calculation, which captures the physics
of bicontinuous topologies. We thus represent a generic bicontinuous
phase as a wedge, shown in Figure~\ref{fig:chip}: an infinitesimal patch of
``saddle'' surface, with edges given by the normals, terminating in a small
line segment lying along the bond-lattice. We envision the surface as
a minimal surface, which has zero mean curvature.\cite{nitsche}

The stretching free energy per molecule of the symmetric wedge
can be calculated as before, given the relative area as a function
of relative height along the center normal (Table~\ref{table1}):
$a(\lambda)\!=\!\lambda(2-\lambda)$, where $\lambda\!=\!z/R$, $z\!=\!0$
is the thin end of the wedge and $z\!=\!R$ is the patch of minimal
surface. As before, the dividing surface location $z_d$ is determined
by eq~\ref{eq:zdloc}.  The resulting free energy is given by applying
eqs~\ref{eq:zdloc}-\ref{eq:interf}, and is shown with the various
cylindrical bounds in Figure~\ref{fig:classicalb}. This estimate misses
by a few tenths of one percent the intersection of the lamellar phase
and the kinked-path upper bound bound for the hexagonal phase, and is
stable with respect to the straight-path upper bound.
{\begin{figure}
\displaywidth\columnwidth
\epsfxsize=3.5truein 
\centerline{\epsfbox{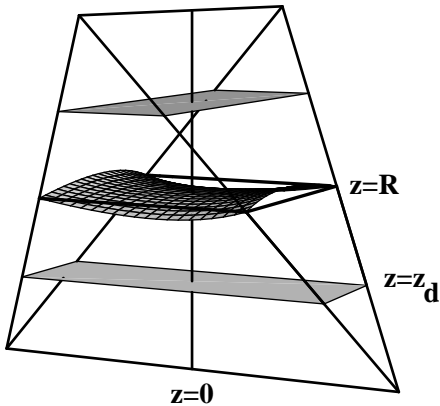}}  
\caption{Symmetric saddle chip ($p\!=\!2$ in Eq.~\ref{eq:p}).
The $A\!-\!B$ dividing surface at
$z\!=\!z_d$ is shaded. $A$ material fills the top and bottom of the
wedge, and $B$ material fills the volumes near the (minimal)
partitioning
surface.}
\label{fig:chip}
\end{figure}}

Clearly the simple wedge construction captures some important physics.
The structures formed by copolymers at different volume fractions $\phi$
arise from competition between interfacial and stretching free energies.
The different structures present different functions $a(\lambda)$, which
determine both the dividing surface area and the stretching energy as a
function of volume fraction. The phases occur in the order they do because
the progression of functions $a(\lambda)$ from quadratic $\lambda^2$
(spheres) to linear $\lambda$ (cylinders) to $\lambda(2-\lambda)$
(bicontinuous) to constant (lamellae) gives progressively less volume to
the ``outer'' chain to avoid stretching, but uses progressively less
area to separate the two species at higher volume fractions of the
minority species.
\begin{figure}
\displaywidth\columnwidth
\epsfxsize=3.5truein 
\centerline{\epsfbox{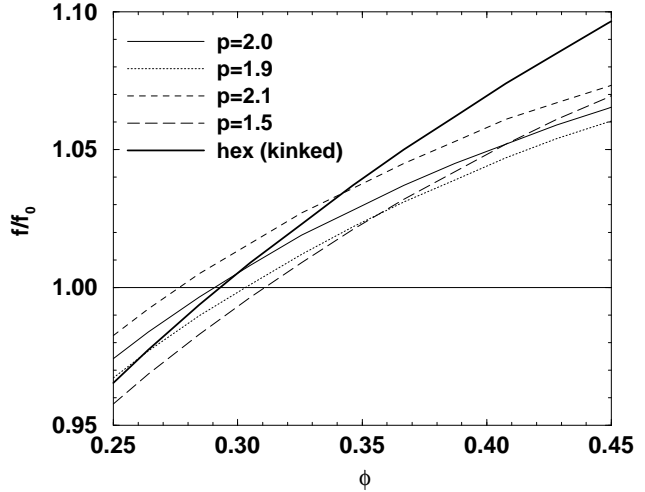}}  
\caption{Free energies for saddle wedges of various shape factors $p=
1.5, 1.9, 2.0, 2.1$. Also shown is the free energy of the kinked-path
hexagonal phase.}
\label{fig:PNwedges}
\end{figure}

However, we cannot argue as before that our simple estimate for the
bicontinuous phase is a lower bound for actual bicontinuous phases,
because there is no way to pack together copies of one infinitesimal
wedge to produce a ``micelle'' that 1) fills some region of space, and
2) is bounded by some surface(s), the volume outside of which could be
filled by homopolymer. Neither can we argue that this estimate is an
upper bound, because we certainly cannot pack a unit cell of the region
bounded by the D or G surfaces with identical copies of one infinitesimal
wedge. 

Bicontinuous phases are assembled from different wedges $\alpha$, with
different shape factors $p_{\alpha}$ in Table~\ref{table1}. The shape
factor roughly gauges the splay or Gaussian curvature of the surface
at the top of the wedge, with $p=1$ (cylinders) corresponding to zero
Gaussian curvature, $p<1$ to positive Gaussian curvature, and $p>1$ to
negative Gaussian curvature. [The Gaussian curvature is the product of
the two radii of curvature of a surface]. The distribution of wedges
must be chosen to pack the desired structure. While the symmetric
wedge has $p=2$, this is not an optimum shape. In fact, the optimum
wedge shape depends on composition, as can be seen in
Figure~\ref{fig:PNwedges}. It is evident that there are shape factors
$p$ which have lower free energy than the straight- and kinked-path
hexagonal upper bounds (Figure \ref{fig:wedgeshape}), so it is not
unreasonable to hope that a judicious packing configuration can be a
stable thermodynamic phase.

The effect of the shape factor $p$ on the topology of the phase
diagram emerges upon examining conformationally asymmetric
($\varepsilon\neq 1$) copolymers. Following Ref.~\cite{Miln94b}, we
explore the effect of conformational asymmetry on the stability of
bicontinuous phases by multiplying the wedge free energy by an
additional arbitrary small prefactor $(0.99)$ which enhances
stability. [We will see below that for $\varepsilon=1$ the
bicontinuous phases that we can calculate (G, D, P) are of order
$0.3-2.2\%$ higher than the cylinder-lamellar crossing, depending on
which upper bound one compares.] Figure~\ref{fig:asymwedge} shows
`phase diagrams' as a function of conformational asymmetry
$\varepsilon$.
		
Recall that for $\varepsilon>1$ the $B$-block is more flexible, while
the $A$-block is stiffer and better able to stretch.  For the symmetric
wedge $(p=2)$ conformational asymmetry reduces the stability of the
stiff-minority wedge phase $(\phi<1/2)$ and enhances the stability of the
flexible-minority wedge phase ($\phi>1/2$), and shifts all transitions to
greater $\phi$. For $p>2.0$ the wedge phases lose stability, as could be
guessed from Figure~\ref{fig:wedgeshape}. For $p<2.0$, for which the wedge
is more cylindrical-like ($p=1$ coresponds to cylinders), conformational
asymmetry enhances the stiff-minority wedge phase relative to both the
lamellar and cylindrical phases, and decreases the stability of the
flexible-minority wedge phase. We emphasize that these are not phase
diagrams, for a true phase is a mixture of wedges with different shape
factors which fill space.
{\begin{figure}
\displaywidth\columnwidth
\epsfxsize=3.5truein 
\centerline{\epsfbox{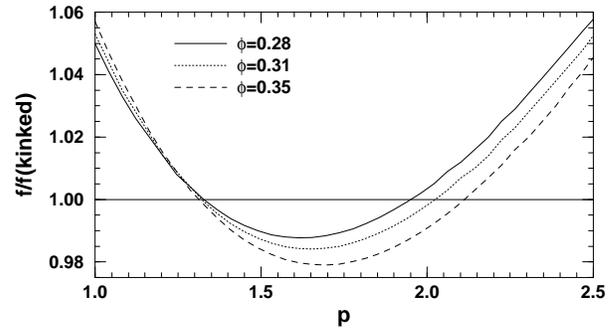}}  
\caption{Free energy relative to that of the kinked-path hexagonal
phase, as a function of saddle wedge shape factor $p$, for compositions
$\phi=0.28, 0.31, 0.35$.}
\label{fig:wedgeshape}
\end{figure}}

\end{multicols}
\widetext 
{\begin{figure}
\displaywidth\columnwidth
\epsfxsize=7.0truein 
\centerline{\epsfbox{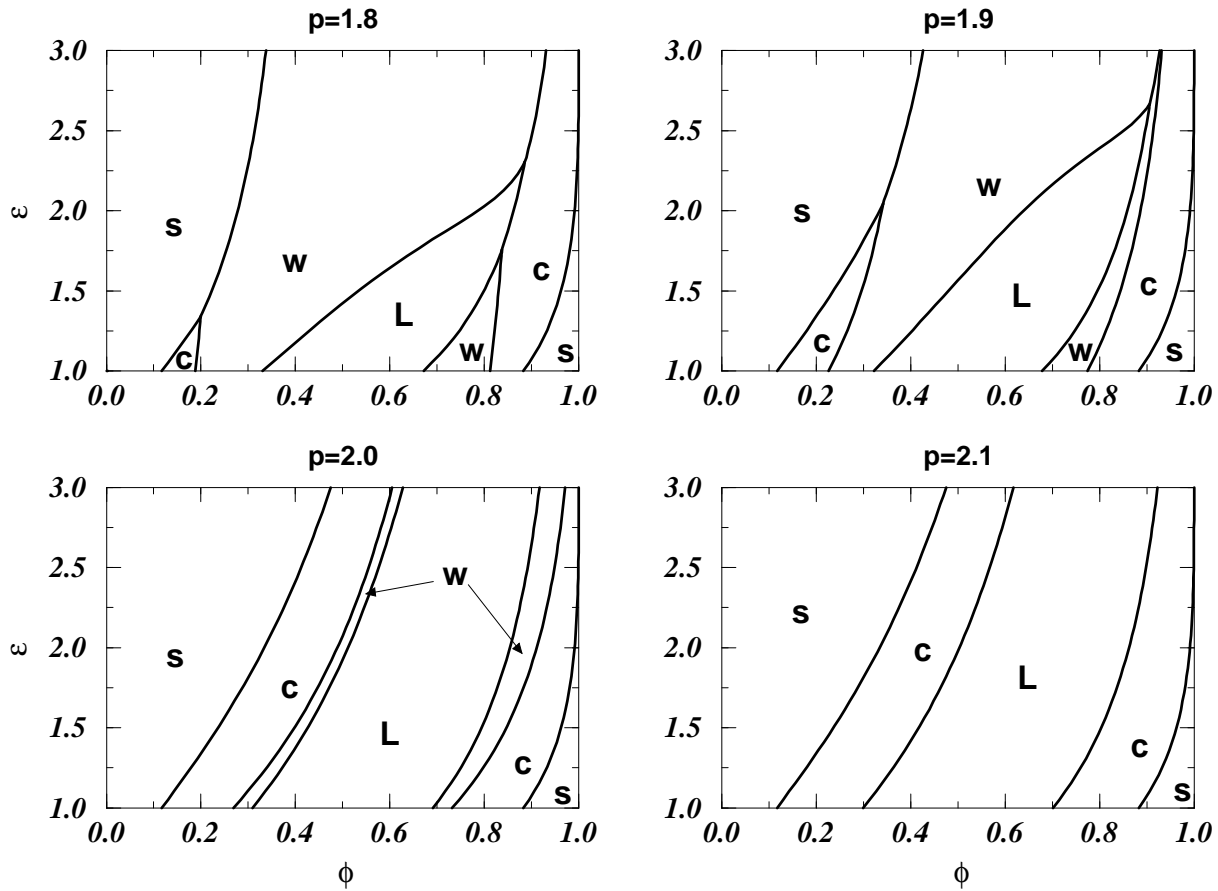}}  
\caption{
Phase diagrams for the round unit cell approximation, including the
saddle wedge for shape factors $p=1.8, 1.9, 2.0, 2.1$, as a function of
composition $\phi$ and conformational asymmetry $\varepsilon$. The free
energy
of the wedge has been multiplied by an additional prefactor $(0.99)$ to
explore the effect of conformational asymmetry on the stability of 
bicontinuous phases.
}
\label{fig:asymwedge}
\end{figure}}
\begin{multicols}{2} 
\narrowtext 

\subsection{Conformational Asymmetry}
Experimentally, the G phase has been observed between the lamellar and
cylindrical phases in several strongly-segregated copolymer systems,
for $\phi$ around 0.3 (and, symmetrically, around 0.7). Some groups
have argued for phase stability of bicontinuous phases in terms of
bending rigidities,\cite{ajdarileibler91,Wang90a}. However, there is
fundamentally no bending energy in the problem; descriptions in terms
of bending energies only arise from a proper accounting of stretching
free energy in curved geometries. Our approach is to choose a geometry
as an ansatz and compute the corresponding interfacial and stretching
free energies in a manner consistent with calculations for cylindrical,
spherical, and lamellar phases. The structure, revealed by scattering
and electron microscopy,\cite{Hajd+94,Schu+96} studies,
can be described as follows,\cite{nitsche,anderson90}.

Consider first the D geometry, which is easier to visualize. 
A skeleton formed of the bonds of a diamond lattice is shown in 
Figure~\ref{fig:skeletons}. Two such lattices interpenetrate,
analogous to the interpenetration of two simple cubic lattices
in a BCC structure. Now imagine swelling the bonds in these
lattices into tubes of a finite diameter. The walls of these tubes
are a rough approximation of the experimentally observed ``dividing
surface'' separating the regions containing the two blocks. The volume
contained within the tubes corresponds to the region inhabited by the
low volume-fraction monomer.

To model the D geometry, we use a self-dual minimal surface,
called the Schwartz~D (diamond) surface, which partitions
space into two identical interpenetrating regions, each of which
contains and is topologically equivalent to a diamond bond-lattice
\cite{thomas86,anderson88,anderson90}. Within each of these regions
is a dividing surface, which surrounds a copy of the bond-lattice. The
copolymer chains then have conformations with one (A) species stretching
towards the bond-lattice, the junction between blocks residing on the
dividing surface, and the other (B) species stretching towards the
minimal surface. In the G phase the diamond lattice is replaced by
a three-fold coordinated lattice, and the surface is replaced by the
gyroid minimal surface discovered by Schoen in 1970 \cite{schoen70},
in which the two interpenetrating G volumes are chiral enantiomers of
one another. For the P phase the bond lattice is six-fold coordinated
(Figure~\ref{fig:skeletons}) and the candidate partitioning surface is
the Schwartz~P minimal surface.

There is no compelling reason to choose a minimal surface for the
partitioning surface. However, minimal surfaces solve the variational
problem of minimizing surface area with zero pressure across the
interface. If the diblock phase is in fact partitioned into two equivalent
connected regions, then by symmetry there can be no net pressure exerted
across the dividing surface that separates the two equivalent disjoint
connected regions.  So a minimal surface is reasonable, but by no means
certain, since there is no obvious area energy to minimize.
{\begin{figure}
\displaywidth\columnwidth
\epsfxsize=2.5truein 
\centerline{\epsfbox{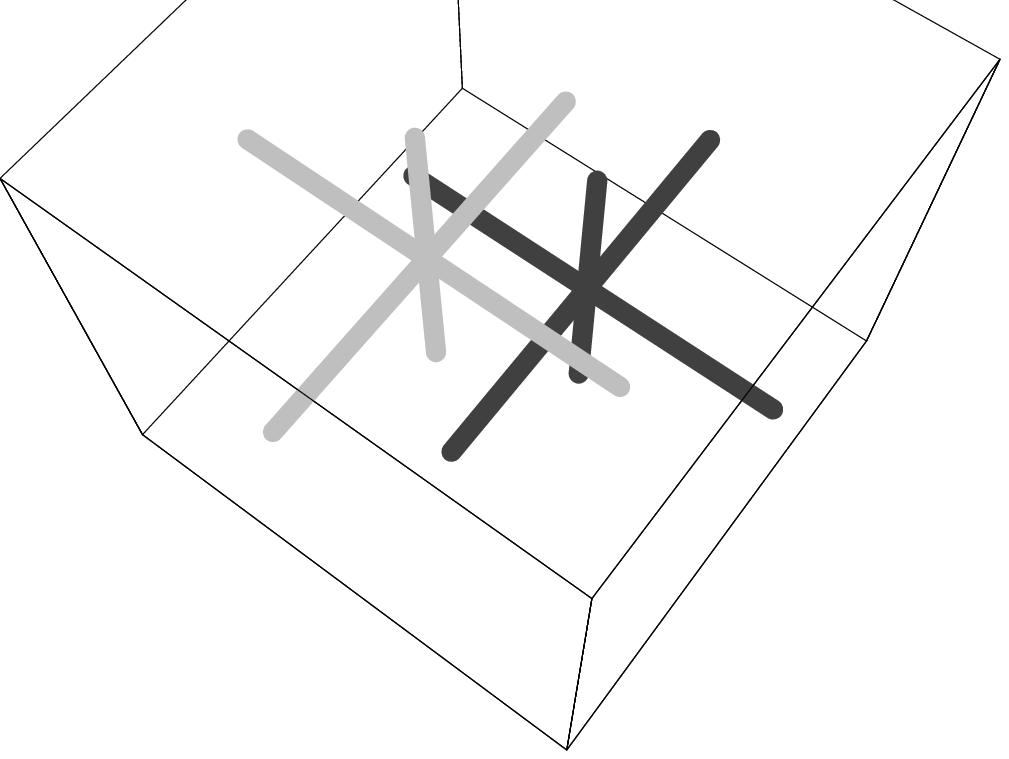}}  
\epsfxsize=2.5truein 
\centerline{\epsfbox{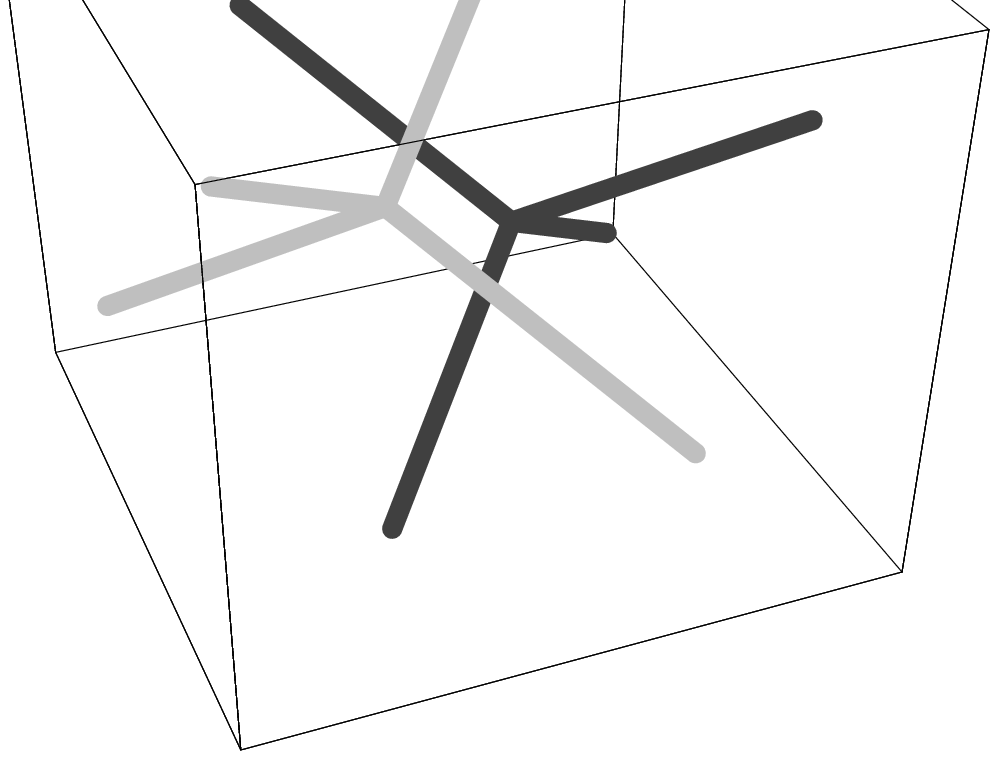}}  
\epsfxsize=2.5truein 
\centerline{\epsfbox{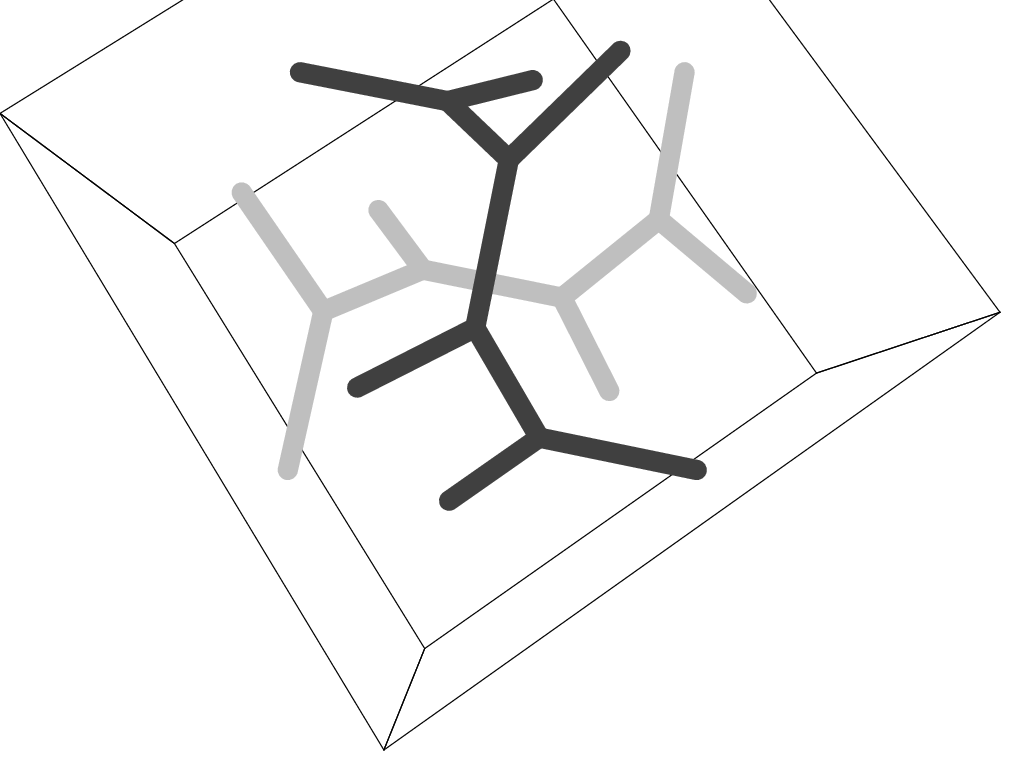}}  
\caption{
Skeleton bond lattices for, from top to bottom, the P, D, and G
surfaces.}
\label{fig:skeletons}
\end{figure}}

The P, D, and G surfaces may be conveniently calculated using the
Weierstrass representation.\cite{nitsche} Here, the three-dimensional
points $\vec{r}$ of the two-dimensional surface are parametrized by the
complex number $\omega=\omega_1 + i \omega_2$. The G, P, and D surfaces
are triply-periodic minimal surfaces with space groups $Ia\bar{3}d,
Im\bar{3}m,$ and $Pn\bar{3}m$, respectively.\cite{schoen70} To generate
the full surface it is enough to calculate a single patch, to which all
the symmetry operations of the space group may be applied to generate
the full structure. A generic surface has two radii of curvature which
are generally non-zero and different. Points where the surface is flat
are  singular points, since both radii of curvature are zero and
a direction of the surface cannot be determined. These flat points define
the corners of the fundamental patch.

The Weierstrass representation is:
\begin{equation}
\vec{r} = \hbox{Re} \int_{{\cal V}} {\vec{v}(\omega)\,e^{i\theta}
\over (\omega^8 - 14 \omega^4 + 1)^{1/2}} d^2\!\omega ,
\end{equation}
where 
\begin{equation}
\vec{v}(\omega) = \left\{ 1-\omega^2, i \left(1 + \omega^2\right),
2\omega\right\},
\end{equation}
and ${\cal V}$ is the domain of integration shown in Figure~\ref{fig:Weier}.
The points on the corners correspond to the flat points, and it is
evident that the integrand above (excluding the measure) is singular
at these points. The angle $\theta$ determines the surface:
\begin{equation}
\theta = \left\{
\begin{array} {c@{\quad}l}
0 & \hbox{D\quad($Pn\bar{3}m$)} \\
90^{\circ} &\hbox{P\quad($Im\bar{3}m$)} \\
38.015^{\circ} &\hbox{G\quad($Ia\bar{3}d$)} 
\end{array} \right.
\end{equation}
{\begin{figure}
\displaywidth\columnwidth
\epsfxsize=2.1truein 
\centerline{\epsfbox{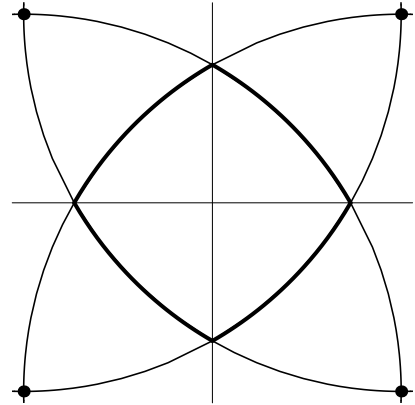}}  
\caption{
The thick line bounds the domain of integration ${\cal V}$ for the
Weierstrass 
representation. The arcs are from circles of radius $\sqrt{2}$ centered
at the points $(\pm 1, \pm1)/\sqrt{2}$.}
\label{fig:Weier}
\end{figure}}

\end{multicols} 
\widetext 
{\begin{figure}
\displaywidth\columnwidth
\epsfxsize=7.0truein 
\centerline{\epsfbox{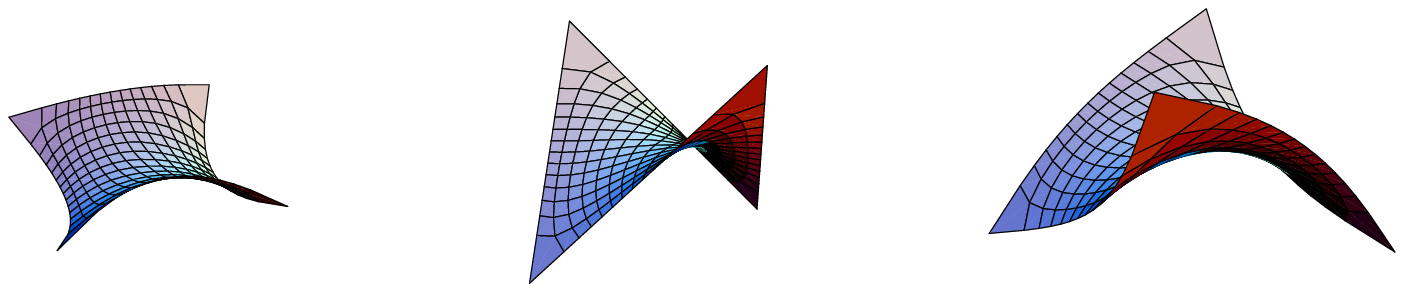}}  
\vskip1.0truecm
\centerline{\epsfbox{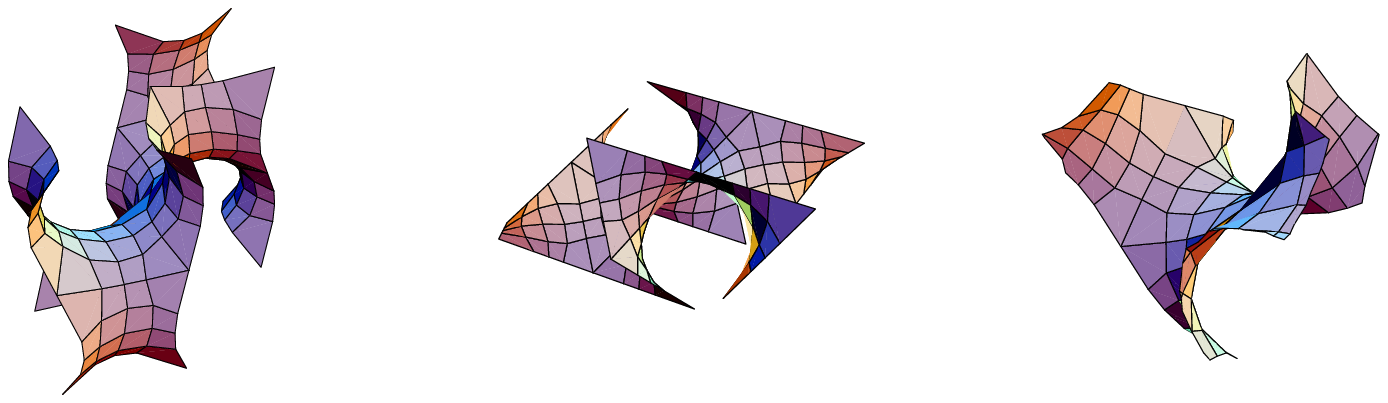}}  
\caption{
Top: fundamental patches of, from left to right, the P, D, and G
surfaces.
Below: larger portions of the surfaces.}
\label{fig:surfaces}
\end{figure}}
\begin{multicols}{2} 
\narrowtext 

For other angles the surface intersects itself. This does not, of
course, exhaust the class of triply periodic minimal surfaces either
mathematically,\cite{schoen70,anderson90,gozdz96b} or physically
\cite{luzzati96}. We have chosen the D, P, and G surfaces because they
are the most common observed surfactant bicontinuous surfaces, and
have been claimed experimentally in block copolymers.  Sections of
these surfaces are shown in Figure~\ref{fig:surfaces} For fairly
accurate calculations (yielding energies \emph{lower} than those for
the true surface by of order a few tenths of a percent) the D surface
may be approximated by a simple hyperbolic surface, $z=xy$. This
suggests that the minimal D surface may not the optimal partitioning
surface.  However, we have varied the shape of the partitioning
surface around the D surface, and found free energy variations of only
a few tenths of a percent.  Because of this, we have not optimized the
free energy with respect to adjustments in the partitioning surfaces.

To produce an upper bound on the free energy of the bicontinuous
phases we follow a procedure analogous to that used for the classical
cylindrical and spherical topologies. Namely, we divide a unit cell of
these structures into a large number of wedges (shown in
Figure~\ref{fig:PGwedges}), similar to Figure  1, but of varying radii and
Gaussian curvature, and average the free energy per molecule using
eq~\ref{eq:avg}-\ref{eq:SA}. Each structure has a different unit
cell (a big wedge) from which the entire structure may be generated by
applying the symmetry operations of the particular space group.
Figure~\ref{fig:Gchip} shows the fundamental cell for the G structure.

This procedure is straightforward. First we calculate the minimal surface,
and then adopt a convenient mapping from the points on this surface to
the skeleton. Essentially, we construct an interpolation between those
high-symmetry points on the minimal surface with normals that project onto
the underlying bond-lattice.  For the $D$ and $G$ phases, we optimize
the mapping from the partitioning surface to the line segments of the
bond lattice to minimize the free energy. We perform this by a conjugate
gradient algorithm that distorts the two dimensional mesh of points on
the surface, and gains of order $1\%$ in energy.
{\begin{figure}
\displaywidth\columnwidth
\epsfxsize=3.5truein 
\centerline{\epsfbox{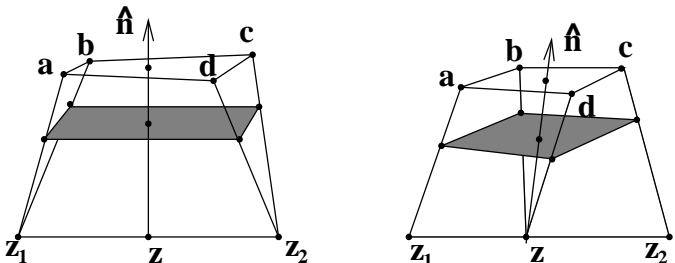}}  
\caption{Fundamental wedges used for constructing the D or P (left) and
G
(right) structures. We have shown quite general wedges, before they
undergo slight adjustments to ensure that ${\bf\hat{n}}$ is
normal to the bond segment $\vec{z}_1\!-\!\vec{z}_2$
and the surface element $\vec{a}\vec{b}\vec{c}\vec{d}$.}
\label{fig:PGwedges}
\end{figure}}
\begin{figure}
\displaywidth\columnwidth
\epsfxsize=3.5truein 
\centerline{\epsfbox{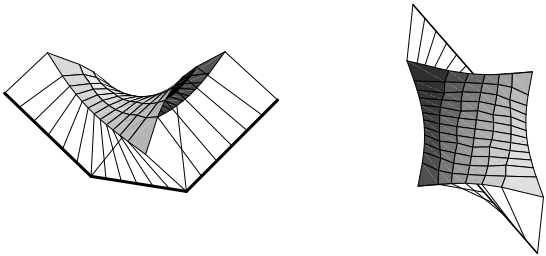}}  
\caption{Side and top views of basic unit for G structure, showing
the initial division into wedges.}
\label{fig:Gchip}
\end{figure}

Thus, each small patch on the minimal surface is connected by straight
lines to a small line segment on the skeleton, and a set of wedges
results.  Each wedge is adjusted slightly by making both the top patch
and the bottom segment orthogonal to the line segment connecting the
center of the patch to the skeleton. This is analogous to making the
outer surface of the wedge in the upper bound for the hexagon phase
orthogonal to the line segment connecting the center of the outer
surface of the wedge to the center of the hexagon. Such adjustments
are negligible in the limit of infinitesimal wedges. As before, the
area of the $A\!-\!B$ interface is the true area, rather than the
(smaller) area that results from adjusting the wedge to assure a chain
path normal to the $A\!-\!B$ interface.

In this way, the unit cell of the region bounded by a minimal surface
is decomposed into many small wedges, each with a known (and different)
radius $R$, shape factor $p$, and volume. The location of the
dividing surface within each wedge is fixed by eq~\ref{eq:zdloc2},
and the free energy calculated with eqs~\ref{eq:avg}-\ref{eq:SA}.
We have checked the algorithm and the dependence on the fineness of
the mesh by using it to successfully compute the free energies of the
hexagonal phase.

\section{Results}\label{sec:results}

{\begin{figure}
\displaywidth\columnwidth
\epsfxsize=3.5truein 
\centerline{\epsfbox{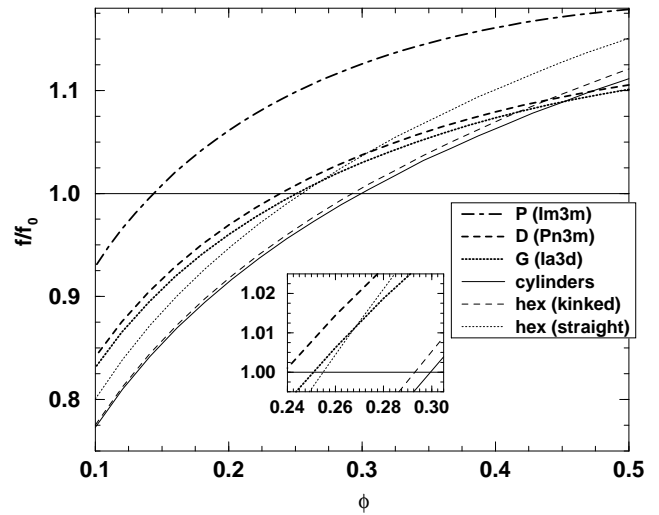}}  
\caption{Free energies for bicontinuous phases (upper bounds) compared
to upper and lower (cylinders) bounds of the hexagonal phase. The inset 
enlarges
the crossing of the G and straight-path-hexagon upper bounds. Asymmetry
parameter $\varepsilon=1.0$.}
\label{fig:data2}
\end{figure}}
Figure~\ref{fig:data2} shows the free energy curves as a function of
$\phi$ for conformationally symmetric copolymers ($\varepsilon=1$).
The upper bound for the $Im\bar{3}m$ phase (P) lies $4-5\%$ above that
for the $Pn\bar{3}m$ phase (D), which in turn is less than a percent
($\simeq 0.7\%$ at $\phi=0.3$) above the $Ia\bar{3}d$ (G) phase.
Consistent with experiments and self-consistent field theory, we do
not find a stable G phase.  At the lamellar--kinked-path-hexagons free
energy crossing ($\phi\simeq 0.293$) the free energy of the $G$ phase
is of order $2.2\%$ larger, while at the
lamellar--straight-path-hexagons free energy crossing ($\phi\simeq
0.255$) the free energy of the $G$ phase is only a few tenths of a
percent ($\simeq 0.28\%$) greater.  This, we have argued, may be the
fairer comparison, since both calculations use straight paths.
Unfortunately, we do not know how to perform a kinked-paths estimate
for the bicontinuous phases.

Note, however, that we do not expect to gain as much energy from a
kinked-path calculation for the bicontinuous phases as for the
cylindrical topologies. Consider the hexagonal calculation. The
boundaries of the Wigner-Seitz cell, and hence the $A\!-\!B$ dividing
surface, have sharp corners into which the chains must stretch.
Presumably a large part of the gain in the kinked-path calculation
comes from relieving the strain associated with this stretch, and
relaxing the inner block to its preferred circular structure.
Bicontinuous phases, on the other hand, have smooth ``Wigner-Seitz
boundaries'' ({\sl i.e.\/} the minimal surface), and expensive
stretching occurs mainly at the junctions of the skeleton lattice.
Hence, the anomalous stretching that may be relieved by a kinked-path
calculation occurs along points in the structure, rather than along
lines. So we expect that our straight-path estimate is not likely to
differ greatly from a kinked-path estimate, and that the free energy
of the G phase remains well above that of the kinked-path hexagons.
The stretching of the chains at the junctions presumably contributes
to the relative stability of the $P$, $D$, and $G$ structures, which
have 6-, 4-, and 3-fold coordinated bond lattices. The more
highly-coordinated lattices require more chain stretching to
accommodate the space, which suggests that $P$, $D$, and $G$ occur in
increasing order of stability. This picture is corroborated by recent
work of Matsen and Bates \cite{matsen96}, who quantitatively examined
the packing frustration in the G, D, and HPL phases.

\end{multicols}
\widetext 
{\begin{figure}
\displaywidth\columnwidth
\epsfxsize=7.0truein 
\centerline{\epsfbox{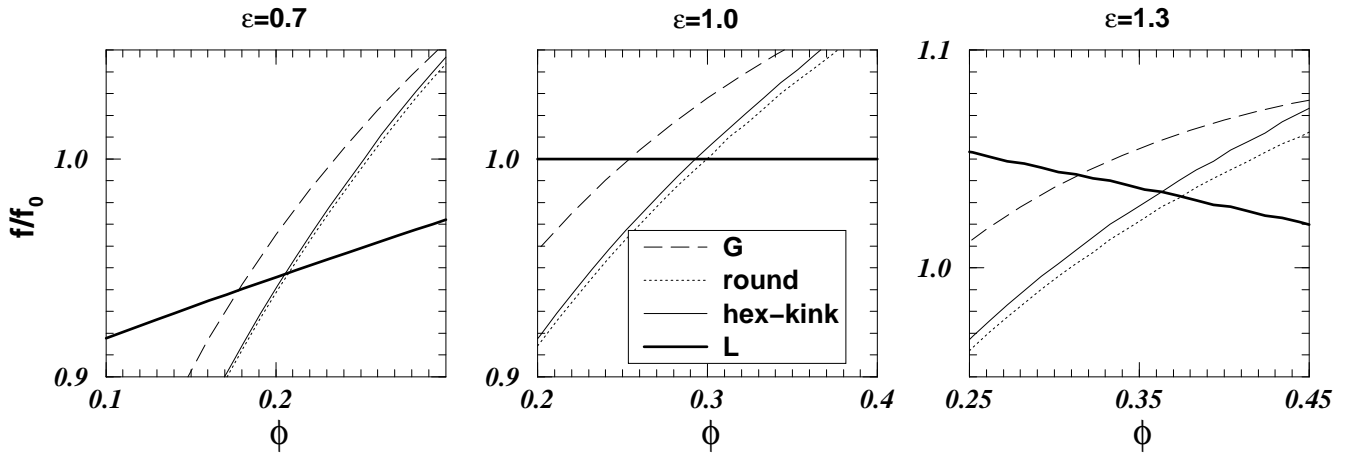}}  
\caption{Free energy crossings for gyroid, hexagonal (kinked paths) 
and lamellar phases for asymmetry parameter values $\varepsilon=
0.7, 1.0, $ and $1.3$. For $\varepsilon>1$ the species 
($B$) with composition $1\!-\!\phi$ is
more flexible and/or has more arms than the other  ($A$) species,
and the $A$ species has an enhanced tendency to stretch.}
\label{fig:asym}
\end{figure}}
\begin{multicols}{2} 
\narrowtext 
Our results apply to the strong-segregation limit, which is attained in
the limit of large $\chi$. Because the phase boundaries shift away from
$\phi=0.5$ as $\chi$ increases from weak segregation \cite{leibler80},
we expect our phase boundaries to be further from $\phi=0.5$ than
experimental values, which is indeed the case. 

Figure~\ref{fig:asym} shows free energy crossings for various values of
the conformational asymmetry parameter $\varepsilon$. Relative to a
conformationally symmetric melt, conformational asymmetry stabilizes
phases with a stiff minority species and destabilizes phases with
a stiff majority species, moving boundaries to larger $\phi(A)$ for
$\varepsilon>1$.  Recall that for $\varepsilon<1$ the inner $A$-block
is more flexible, while the $B$-block is stiffer and better able to
stretch.  We find that $\varepsilon<1$ reduces the relative stability
of the G phase, while the stability is enhanced for $\varepsilon>1$,
and becomes stable for rather large asymmetries $\varepsilon\agt 9.0$
(Figure~\ref{fig:asymG}).

{\begin{figure}
\displaywidth\columnwidth
\epsfxsize=3.5truein 
\centerline{\epsfbox{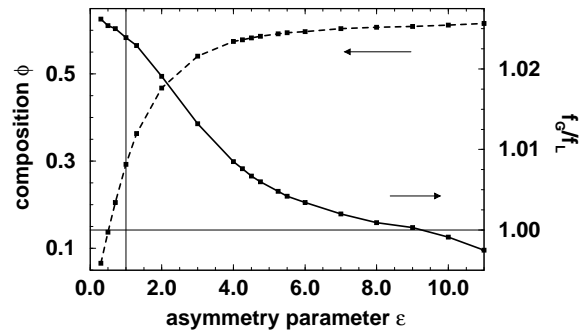}}  
\caption{Free energy of G phase relative to that of the kinked-path
upper bound estimate of the lamellar-hexagonal phase boundary,
as a function of conformational asymmetry $\varepsilon$. The vertical
line
marks the conformationally symmetric copolymer, and the horizontal line
is for
reference.}
\label{fig:asymG}
\end{figure}}

Previous calculations of phase diagrams of
conformationally-asymmetric diblocks have been done in the weak
segregation regime,\cite{vavasour93,matsen97a} and, for the generic
symmetric wedge  as a model bicontinuous structure, in the strong
segregation regime.\cite{Miln94b} Matsen and Bates \cite{matsen97a}
found, as we do, that conformational asymmetry stabilizes the G phase
with a stiff minority phase, widening the composition window and moving
the lamellar-G and G-cylinder boundaries to higher stiff compositions;
and destabilizes the G phase with a stiff majority phase, both narrowing
the composition window and shifting it to a higher stiff fraction.
Their calculations are limited to $\chi N\alt 30$, and it is inconclusive
whether the $\chi N \rightarrow\infty$ limit of this calculation yields a
stable G phase.  The same qualitative behavior was found for the generic
symmetric wedge in the strong segregation regime.\cite{Miln94b}

Table~\ref{table2} summarizes asymmetry parameters from recently
collected data \cite{fetters94,faraday}. While none of these diblocks
have the large conformational asymmetry required to test our
prediction of a stable phase in strong segregation, $A_n-B_m$
starblock copolymers have asymmetry factors larger, by a factor of
$n/m$, than those due to intrinsic chain stiffness effects alone. For
example a $\hbox{PE}_6$--$\hbox{PEE}_1$ starpolymer has an asymmetry
factor $\varepsilon\simeq 6\times 1.58 =9.48$.

We have attempted to calculate energies for the HPL
\cite{hamley93,disko93} phase, which is now accepted as a metastable phase
\cite{hajdukunpub}. The minimal crystal phases D, P, and G have obvious
candidate minimal surfaces to act as an intermaterial dividing surface
towards which the majority-phase ends stretch; and the minority-phase
ends stretch towards the skeletal bond lattice. On the other hand, the
majority-phase ends in the HPL phase stretch towards a combination of
lines (in the hexagonally-arranged perforating tubes) and surfaces (within
the majority-phase layer). Similarly, it is not obvious how to partition
the minority-phase ends between lines and surfaces. The result is a
non-analytic mapping which is difficult to minimize over. Our attempts
have thus far yielded quite high energies, of order that of the $P$ phase.

\section{Summary}\label{sec:summary}
We have outlined a general method for computing the free energy of
block copolymer phases in the strong segregation regime. The procedure
consists of the following steps: 
\begin{enumerate} 
\item Choose a candidate geometry and an associated partitioning surface
that divides space into disjoint interpenetrating regions (the majority
blocks from the two regions stretch towards this surface).
\item  Divide the enclosed volume into infinitesimal wedges, defined by
straight paths connecting the partitioning surface to a skeleton of
bonds (the minority blocks stretch towards this bond skeleton).
\item The A-B interface in each wedge is located such that the fraction
of wedge volume filled by A blocks is locally equal to $\phi$. The
interfacial contribution to the free energy is the area of the A-B
interface times the A-B surface tension.
\item Compute the stretching free energy per chain for each wedge within
the approximation of straight paths.
Calculations for straight paths involve
slight adjustments to the wedges whose contributions vanish in the limit
of small wedges.  
\item Optimize the free energy per chain with respect to the overall
scale of the mesophase ({\sl e.g.\/}, the dimension of the unit cell).
\item Optimize the mapping from the partitioning surface to
the bond skeleton to minimize the overall free energy.  
\end{enumerate}
In certain structures ({\sl e.g.\/} HPL) the majority phase ends lie
on both lines and surfaces, in which case the procedure above must be
suitably generalized. We have also shown how to calculate the free
energy for geometries where the shape of the $A\!-\!B$
interface is specified, for phases of cylindrical topology. This
requires polymer chain paths which are kinked at the $A\!-\!B$ interface.

The infinitesimal wedges are described by the relative area function
$a(z/R)$ (eq~\ref{eq:area}) which is parametrized by a single scalar $p$
(eq~\ref{eq:p} and Table~\ref{table1}) that roughly gauges the local
Gaussian curvature of the partitioning surface. For the classical phases
all wedges are identical, while bicontinuous phases have different
distributions of shape factors $p$. For symmetric stars we find a
metastable bicontinuous (gyroid, or G) phase which is most stable near
the lamellae-hexagonal cylinder transition.  For sufficiently asymmetric
copolymers ($\varepsilon\agt 9.0$) we predict a {\sl stable\/} G phase.


\begin{thebibliography}{10}
\bibitem{mmnote} The composition variable $\phi$ is defined in
Ref.~\cite{Miln94b} as the volume fraction of the $B$ phase. Similarly,
Ref.~\cite{Miln94b} defines $\varepsilon$ with $A$ and $B$
reversed with respect to the present article, so equations such
as eqs~\ref{eq:felam}-\ref{eq:fesph} are identical in the two works
(compare eqs~9-11 of Ref.~\cite{Miln94b}).  Eq~8 of Ref.~\cite{Miln94b}
has a misprint, and should have a factor of $(n_{\scriptscriptstyle
A} n_{\scriptscriptstyle B})^{1/3}$ to match eq~\ref{eq:f0} in
Section~\ref{sec:general} above.
\end{thebibliography}

\end{multicols} 
\end{document}